\documentclass[a4paper,10pt]{article}
\usepackage[utf8]{inputenc}
\usepackage{amsthm}
\usepackage{amssymb}
\usepackage{amsmath}
\usepackage{graphicx}
\usepackage{colortbl}
\usepackage{color}
\usepackage{xcolor}
\usepackage{textcomp}
\usepackage{subcaption}
\begin{document}
\begin{center}
{\Large \textbf{\centering High resolution $\gamma$-ray spectroscopy using GALILEO array}}
\end{center}

 \vspace*{0.5cm}

D.~Testov$^{1,a,*}$, J.J.~Valiente-Dob\'on$^2$, D.~Mengoni$^1$, F.~Recchia$^1$, A.~Goasduff$^1$, A.~Boso$^{1,4}$, S.~M.~Lenzi$^{1}$, G.~de~Angelis$^2$, S.~Bakes$^{2,b}$, C.~Boiano$^{3}$, B.~Cederwall$^{5}$,  G.~Colucci$^{1}$, M.~Cicerchia$^{2}$,  P.~\v{C}olovi\'{c}$^{6}$, F.~Didierjean$^{7}$, M.~Doncel$^{8}$,  J.A.~Due\~{n}as$^{9}$, F.~Galtarossa$^2$, A.~Gozzelino$^2$,  K.~Hady\'{n}ska-Kl\c{e}k$^2$, R.~Isocrate$^2$, G.~Jaworski$^2$, P.R.~John$^{1, 10}$,  H.~Liu$^{5}$, S.~Lunardi$^{1}$, R.~Menegazzo$^{1}$, A.~Mentana$^{3}$,  V.~Modamio$^{11}$,  A.~Nannini$^{12}$,  D.R.~Napoli$^{2}$, M.~Palacz$^{13}$,  G.~Pasqualato$^{1}$, M.~Rocchini$^{12}$,  S.~Riccetto$^{2,14}$, B.~Saygi$^{2, 15}$, E.~Sahin$^{11}$,  M.~Siciliano$^{2}$,  Yu.~Sobolev$^{16}$, S.~Szilner$^{6}$

\vspace*{1cm}
\textit{\centering
$^{1}$ Dipartimento di Fisica e Astronomia and INFN, Sezione di Padova, Padova, Italy. \\
$^{2}$ INFN, Laboratori Nazionali di Legnaro, Legnaro, Italy. \\
$^{3}$ Dipartimento di Fisica dell’Universit\`{a} di Milano e Sezione INFN, Milano, Italy.\\
$^{4}$ National Physical Laboratory, Teddington, Middlesex, UK \\
$^{5}$ KTH Royal Institute of Technology, Stockholm, Sweden \\
$^{6}$ Ru{d\llap{\raise 1.22ex\hbox
  {\vrule height 0.09ex width 0.2em}}\rlap{\raise 1.22ex\hbox
  {\vrule height 0.09ex width 0.06em}}}er
  Bo\v{s}kovi\'{c} Institute, Zagreb, Croatia.\\
$^{7}$ Institut Pluridisciplinaire Hubert CURIEN (IPHC), Strasbourg, France. \\
$^{8}$ The University of Liverpool, UK \\
$^{9}$ Departamento de Ingenier\'{i}a Eléctrica, Universidad de Huelva, Spain. \\
$^{10}$ Institut f\"{u}r Kernphysik,  Technische Universit\"{a}t Darmstadt, Germany.  \\
$^{11}$ University of Oslo \\
$^{12}$ Universit\`{a} degli Studi and INFN Sezione di Firenze, Florence, Italy.\\
$^{13}$ Heavy Ion Laboratory, University of Warsaw, Poland. \\
$^{14}$ Dipartimento di Fisica e Geologia dell'Università di Perugia e INFN Sezione di Perugia, Perugia, Italy.\\
$^{15}$ Department of Physics, Ege University, Faculty of Science, Izmir, Turkey. \\
$^{16}$ Joint Institute for Nuclear Research, Dubna, Moscow region, Russia\\
$^{a}$ also at Joint Institute for Nuclear Research, Dubna, Moscow region, Russia\\
$^{b}$ also at Department of Physics, University of Surrey. Guildford UK\\
$*$E-mail: testov@lnl.infn.it\\
}

\begin{abstract}
The GALILEO $\gamma$-ray spectrometer has been constructed at the Legnaro National Laboratory of INFN (LNL-INFN). It can be coupled to advanced ancillary devices which allows  nuclear structure studies employing the variety of in-beam $\gamma$-ray spectroscopy methods. Such studies benefit from reactions induced by the intense stable beams delivered by the Tandem-ALPI-PIAVE accelerator complex and by the radioactive beams which will be provided by the SPES facility. In this paper we outline two experiments performed within the experimental campaign at GALILEO coupled to the EUCLIDES Si-ball and the Neutron Wall array. The first one was aimed at spectroscopic studies in A=31 mirror nuclei and the second one at measurements of lifetimes of excited states in nuclei in the vicinity of $^{100}$Sn. 
%
\end{abstract}


\section{Introduction}\label{sec:introduction}

\begin{figure}\centering
\includegraphics[width=3.0in, page=2]{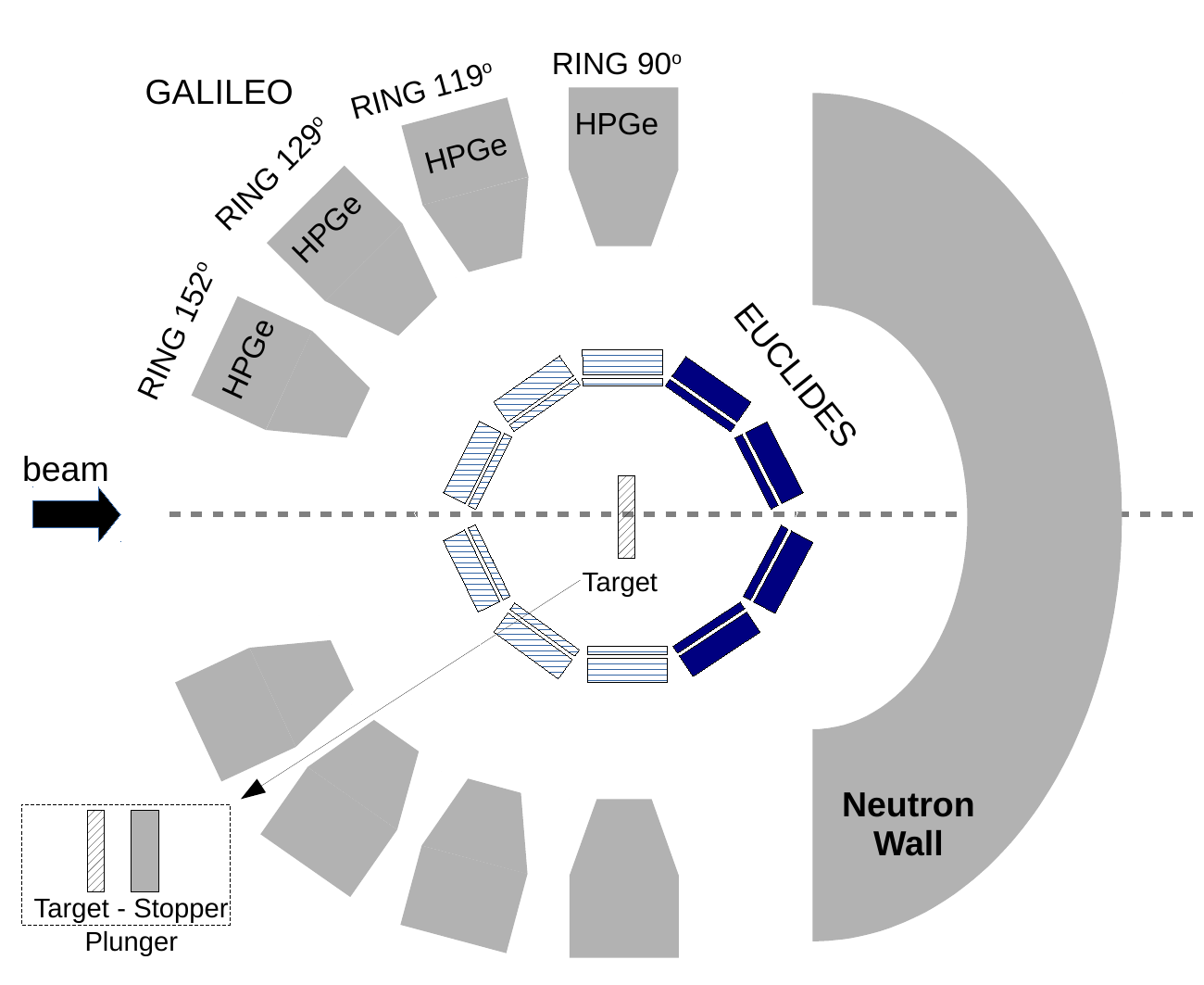}
\caption{Schematical view of GALILEO $\gamma$-ray array coupled to EUCLIDES, Neutron Wall. The target is installed at the centre of the reaction chamber. The hatched part of ECULIDES is dismounted to allow plunger installation. See text for more details.}
\label{fig:setup}
\end{figure}

GALILEO $\gamma$-ray array spectrometer is the resident array at National Legnaro Laboratories. In Phase~I~\cite{Val2014} it consists of 25 Compton-suppressed HPGe tapered detectors, originally from the GASP array~\cite{Baz1992_gasp}.It is organized in 4 rings. Three backward rings made of 5 detectors each at 152$^\circ$, 129$^\circ$ and 119$^\circ$. The last ring at 90$^\circ$ comprises 10 detectors, see Fig.~\ref{fig:setup}. 

The measured absolute efficiency measured is $\sim$2.3\% for 1.3~MeV $\gamma$-ray and the average resolution around 2.5~keV \cite{Zanon_theis}. The Peak to Total ratio is $\sim$50\%. GALILEO can be run in a stand-alone mode or coupled to ancillary devices. During the first experimental campaigns we used one or more ancillary detectors such as the light-charged-particle detector array EUCLIDES~\cite{Tes2016}; pixel-type silicon detector of TRACE~\cite{Men2014, Cie2018}; a heavy ion detector for Coulomb excitation measurements SPIDER~\cite{Roc2016}; the plunger device \cite{Mul2019} for electromagnetic-moment measurements, the NEUTRON WALL~\cite{Ske1999, Lju2004}. To increase the $\gamma$-ray efficiency for high-energy transitions a LaBr$_{3}$ array \cite{Gia2013, LaBr} can be used complementary to other ancillary detectors. The GALILEO electronic system is fully digital and it is synchronized by a distributed clock delivered by the GTS (Global Trigger and Synchronization) system~\cite{Bel2013}, which enables the time synchronization between the GALILEO spectrometer and all other ancillary detectors. Some of the results of this first experimental campaign were already reported in the publications \cite{Hua2016, Zha2018, Qia2019} and in the LNL Annual reports \cite{ar_lnl}. 

The EUCLIDES Si-ball array can be installed inside the chamber in the full or in the plunger configuration. In the full configuration EUCLIDES includes 40~$\Delta$E-E telescopes covering $\sim$80\% of the solid angle.  The light charged particle identification of EUCLIDES relies on the $\Delta$E-E method. Almost 4$\pi$ coverage of the solid angle and high granularity of EUCLIDES ensures its high particle detection efficiency and the possibility to reduce the Doppler broadening of peaks in the recorded $\gamma$-ray spectra by an event-by-event kinematic reconstruction of the trajectory of recoiling nuclei~\cite{Tes2016}. An EUCLIDES configuration with only forward positioned $\Delta$E-E telescopes allows plunger installation as is described in Sec.~\ref{sec:lifetime}. 

\begin{figure}\centering
\includegraphics[width=3.5in]{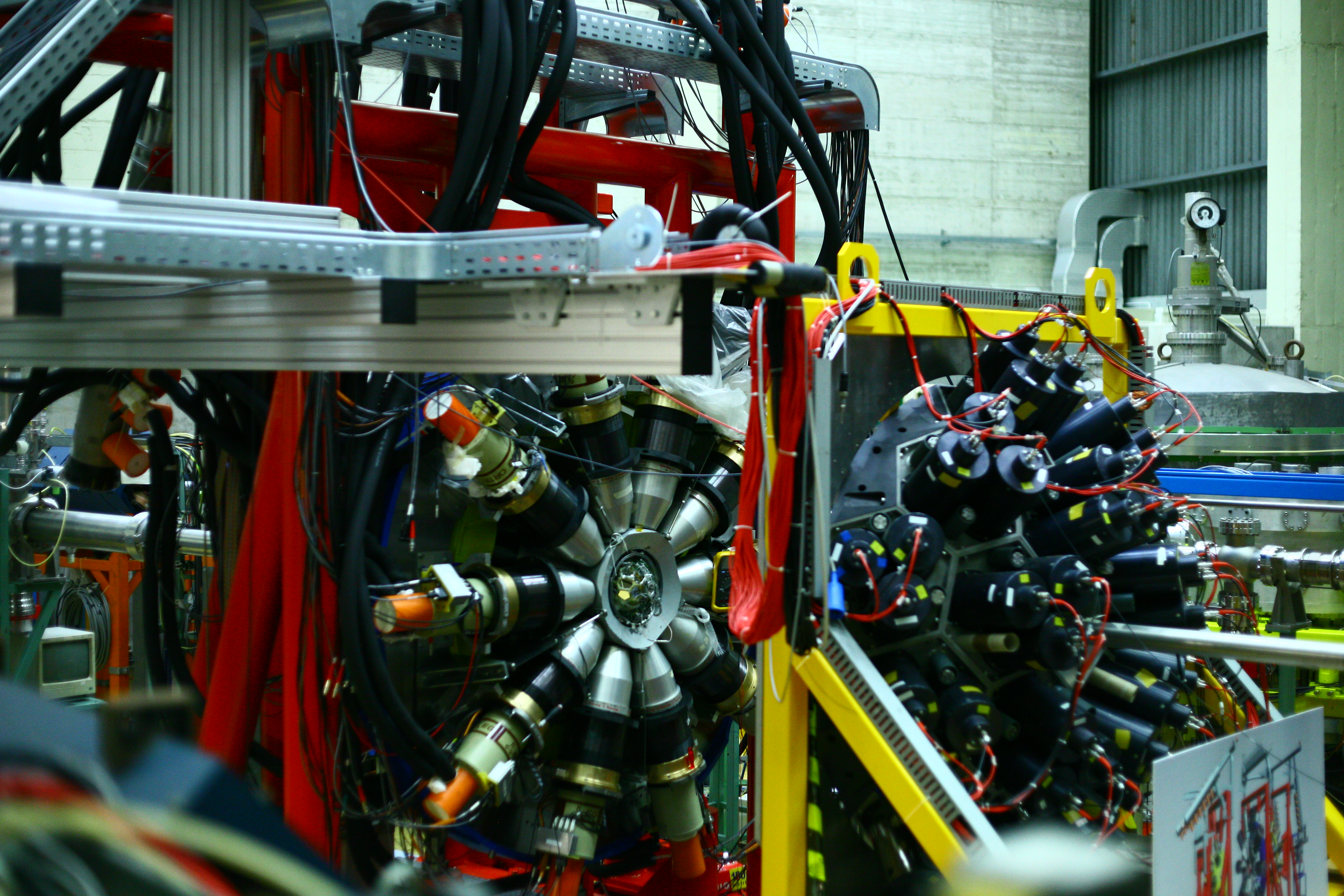}
\caption{GALILEO $\gamma$-ray spectrometer. On the left part it is visible 10 HPGe detectors inside BGO shield placed at 90$^{o}$ around the reaction chamber. Neutron Wall array, positioned on forward angles with respect to the beam direction is visible on the right side. The EUCLIDES Si-ball is inserted in the reaction chamber of GALILEO which can be seen opened at the centre.}
\label{fig:galileo}
\end{figure}

The Neutron Wall, composed of 15 hexagonal detectors arranged in 2 rings around the central pentagonal unit,  is installed on the forwards angles with respect to the beam direction. Each hexagonal unit is divided in 3 hermetically separated segments. Considering the solid angle coverage of $\sim$1$\pi$ the efficiency reached for a single neutron detection is 20-25\%. Performance of Neutron Wall coupled to GALILEO as well as details of pulse-shape-analysis are reported in Ref.~\cite{Lon2015}. The picture of the setup is given in Fig~\ref{fig:galileo}.

To highlight the performance of GALILEO we have selected two experiments. Thus, in Sec.~\ref{sec:a31} it is reported an experiment to investigate excited levels of the $^{31}$S populated in a fusion evaporation reaction. In Sec.~\ref{sec:lifetime} we describe the  measurements of lifetimes of nuclear states in nuclei located in the vicinity  of $^{100}$Sn.

%

\section{High spin states in mirror nuclei $^{31}$S and $^{31}$P}\label{sec:a31}

One of the first experiments performed using GALILEO was aimed at identification of high spin states in mirror nuclei $^{31}$S and $^{31}$P. $^{12}$C beam at 45~MeV impinged on the self-supported $^{24}$Mg target of 400~$\mu$g/cm$^{2}$. The most recent studies of $^{31}$S performed\cite{Jen2005} revealed large oscillation behaviour of MED values for the negative-parity sequence as a function of spin, see Ref.~ \cite{Jen2005}. These oscillation may be explained including in the wave function excitations to the fp shell considering thus the electromagnetic spin-orbit effect. Description of the MED in \textit{sd} shell nuclei for negative parity and high spin states involving the electromagnetic spin orbit term is up to now only qualitative (because it involves interactions in two main shells). Additionally, shell-model calculations performed using the USD residual interaction and the Monte Carlo shell model with the SDPF-M interaction reproduce well the excitation energies and the reduced transition probabilities for positive-parity states up to the spin $\frac{13}{2}^{-}$, see Ref.~\cite{Ion2006}. An interesting feature revealed by these calculations is that the yrast negative-parity states show an alternating structure: the $\frac{7}{2}^{-}$ , $\frac{11}{2}^{-}$ , and $\frac{15}{2}^{-}$ states are described by almost equal contributions of the proton and neutron excitation to the fp shell, whereas the $\frac{9}{2}^{-}$ and $\frac{13}{2}^{-}$ states have only a neutron excitation to the f$_{7/2}$ shell.
On the experimental side, MED values are available up to spin J=13/2 for both negative and positive parity \cite{Jen2005} which is not sufficient to disentangle the theoretical puzzle. Therefore, data on MED values for higher spin states are needed. 

Excited levels of $^{31}$P was were previously studied in $^{12}$C($^{20}$Ne,p(n)), see Ref.~\cite{Jen2005} and $^{24}$Mg($^{16}$O,2$\alpha$p(n)) reactions, Ref~\cite{Ion2006, Ved2005} up to high spins. In contrast, the more exotic $^{31}$S  was observed up to only  $\frac{13}{2}^{+}$ and $\frac{13}{2}^{-}$ spin Ref.~\cite{Jen2005}. Therefore, the goal of reported experiment was to extend the level schemes and study the mirror energy differences in the A=31, T=1/2 mirror nuclei. $^{31}$P and $^{31}$S  were produced in the same fusion evaporation reaction $^{24}$Mg($^{12}$C,$\alpha$p) and $^{24}$Mg($^{12}$C,$\alpha$n) respectively. 

The identification of $^{31}$P (1$\alpha$1p) and $^{31}$S (1$\alpha$1n) was performed in the off-line analysis using EUCLIDES and Neutron Wall. In the preliminary data analysis the sum of projections of $\gamma-\gamma$ matrix recorded requesting coincidence with 1$\alpha$ and 1 neutron is shown in Fig.~\ref{fig:spectra}. The $\gamma$-ray transitions in $^{31}$S known from experiments cited in the literature are marked. A further analysis is ongoing~\cite{Tes2018_mirror}. 

\begin{figure}
\includegraphics[width=3.5in]{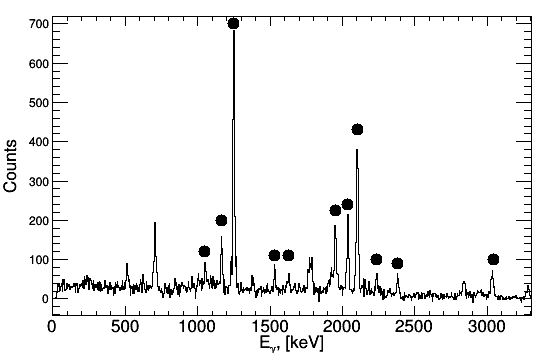}
\caption{Sum of the projections of $\gamma-\gamma$ matrix gated on 1249-keV, 2102-keV and 1166-keV transitions in $^{31}$S.  1$\alpha$ and 1 neutron conditions were requested. The marked peaks correspond to previously known transitions in $^{31}$S.}
\label{fig:spectra}
\end{figure}

\section{Probing the strength of the $^{100}$Sn shell closure via life-time measurements in light Sb and Te}\label{sec:lifetime}

\begin{figure}\centering
\includegraphics[width=3.5in]{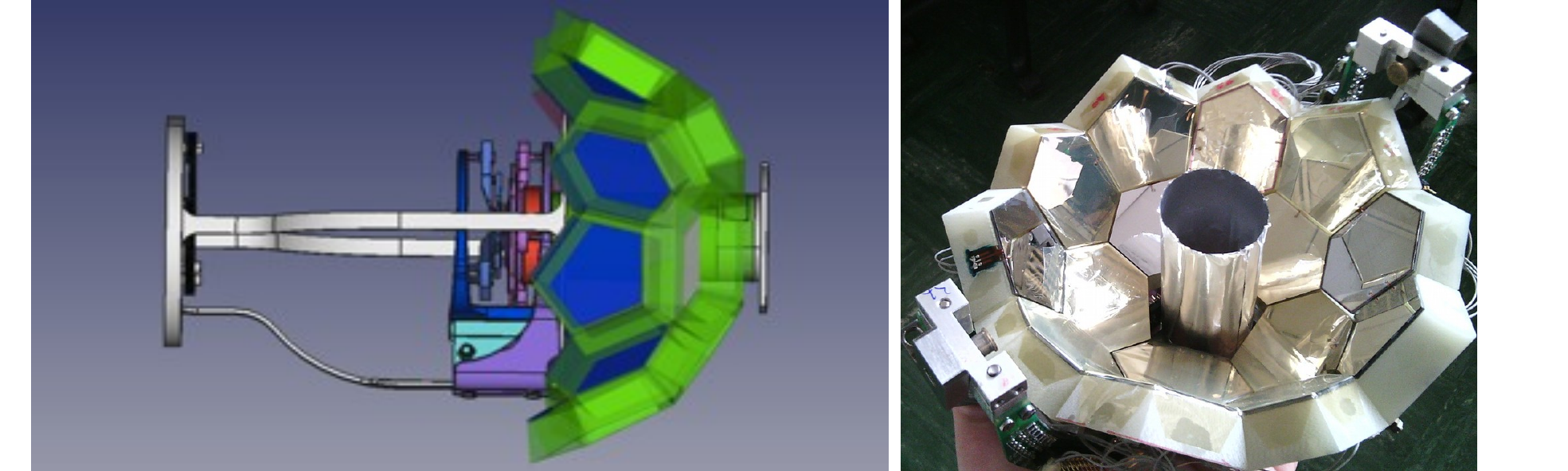}
\caption{Plunger device at LNL coupled to EUCLIDES (left); EUCLIDES plunger configuration consisting of 5 segmented and 10 single-plate $\Delta$E-E telescopes (right).}
\label{fig:plunger}
\end{figure}

The structure of nuclei far from $\beta$ stability has been a key subject for both experimental research and theoretically investigations. The information on the electromagnetic properties of the neutron-deficient nuclei in the vicinity of $^{100}$Sn represent unique testing ground for many different theoretical approaches. Recoil Distance Doppler-Shift (RDDS) method\cite{Dew2012} is extensively used in the nuclear structure experiments to determine excited states lifetime in the range of the ps to the ns. Thus a dedicated plunger device\cite{Mul2019} to fit in the GALILEO reaction chamber with part of the EUCLIDES array was constructed. This allows the application of the RDDS method relying on $\gamma-\gamma$ coincidence measurements. In order to select evaporation channels in a fusion-evaporation reaction we removed  the backward positioned Si-telescopes of EUCLIDES Si-ball array\cite{Tes2016}, see Fig.~\ref{fig:setup}. The removed telescopes contribute weakly to the overall detection efficiency of light charged particles in the present experiment. In this configuration, presented in Fig.~ \ref{fig:plunger}, EUCLIDES consists of 5 segmented $\Delta$E-E telescopes placed at the forward angle and 10 single-plate telescopes in the second forward ring. Thus, the channel selection capability of EUCLIDES can be exploited also in RDDS experiments. 


During the performed experiment neutron-deficient nuclei were populated using a 2~pnA beam of $^{58}$Ni impinged into a 1~mg/cm$^{2}$ $^{58}$Ni target followed by a 15~mg/cm$^{2}$ Au-stopper foil. The quality of the obtained data can be seen in Fig.~\ref{fig:spectra}. In this figure it is shown the coincidence spectra resulting from gating on the 4$^{+}\rightarrow$2$^{+}$ shifted component in $^{112}$Te observed in detector-ring 1 ($\Theta_{1}$=129$^{o}$) in coincidence with 3 protons identified by EUCLIDES. The stopped and the shifted component of 2$^{+}\rightarrow$0$^{+}$ transition in $^{112}$Te are shown for the a set of the target-to-stopper distances. The measured lifetime of the 2$^{+}$ coincide with the previous values cited in the literature and will be reported in a separate publication. The analysis to study the excited states is ongoing\cite{Bak2017}.


\begin{figure}\centering
\includegraphics[scale=0.48]{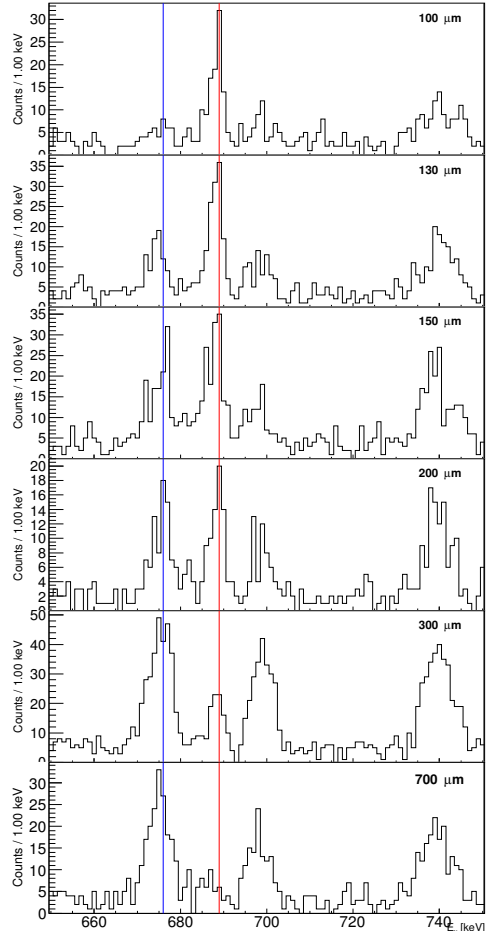}
\caption{Coincidence spectra resulting from gating on the 4$^{+}\rightarrow$2$^{+}$ shifted component in $^{112}$Te observed in detector-ring 1 ($\Theta_{1}$=129$^{o}$). 3p conditioned was requested. Indicated is the 2$^{+}\rightarrow$0$^{+}$ shifted and stopped component in $^{112}$Te in detector-ring 2 ($\Theta_{1}$=119$^{o}$).}

\label{fig:spectra}
\end{figure}

\section{Conclusion}\label{sec:conclusion}

In the paper we briefly describe the new GALILEO $\gamma$-ray spectrometer constructed at LNL Legnaro. It can be coupled to ancillary devices to allow variety of nuclear structure research. In the paper we report preliminary results of two experiments performed using GALILEO during the first experimental campaign at LNL Legnaro. One of the experiments was aimed at spectroscopic studies of high-level states in $^{31}$S, and the second one, to measure lifetime in the nuclei in the vicinity of $^{100}$Sn using the coincidence RDDS method conditioned  on charged particles detected by EUCLIDES array. In the nearest future nuclear structure research at GALILEO will benefit from accelerated beams of radioactive isotopes to be delivered by SPES facility\cite{Bis2016_spes}.

\bibliographystyle{ws-procs9x6} 


\end{document}